\documentclass[nohyperref]{article}

\usepackage{microtype}
\usepackage{graphicx}
\usepackage{subfigure}
\usepackage{booktabs} % for professional tables

\usepackage{hyperref}
\usepackage{multirow}

\usepackage[accepted]{icml2022}

\usepackage{amsmath}
\usepackage{amssymb}
\usepackage{mathtools}
\usepackage{amsthm}

\usepackage{microtype}
\usepackage{placeins}

\usepackage[capitalize,noabbrev]{cleveref}

\theoremstyle{plain}

\theoremstyle{definition}

\theoremstyle{remark}

\usepackage[textsize=tiny]{todonotes}

\icmltitlerunning{DrumGAN VST: A Plugin for Drum Sound Analysis/Synthesis with GANs}

\begin{document}

\twocolumn[
\icmltitle{DrumGAN VST: A Plugin for Drum Sound Analysis/Synthesis with Autoencoding Generative Adversarial Networks}




\begin{icmlauthorlist}
\icmlauthor{Javier Nistal}{yyy}
\icmlauthor{Cyran Aouameur}{yyy}
\icmlauthor{Ithan Velarde}{comp}
\icmlauthor{Stefan Lattner}{yyy}

\end{icmlauthorlist}

\icmlaffiliation{yyy}{Sony Computer Science Laboratories, Paris}
\icmlaffiliation{comp}{National Institute of Applied Sciences, Lyon}

\icmlcorrespondingauthor{Javier Nistal}{j.nistalhurle@gmail.com}
% \icmlcorrespondingauthor{Cyran Aouameur}{first2.last2@www.uk}
% \icmlcorrespondingauthor{Ithan Velarde}{first2.last2@www.uk}
% \icmlcorrespondingauthor{Stefan Lattner}{first2.last2@www.uk}

\icmlkeywords{Machine Learning, ICML}

\vskip 0.3in
]



\printAffiliationsAndNotice{}  % leave blank if no need to mention equal contribution

\begin{abstract}
In contemporary popular music production, drum sound design is commonly performed by cumbersome browsing and processing of pre-recorded samples in sound libraries. One can also use specialized synthesis hardware, typically controlled through low-level, musically meaningless parameters. Today, the field of Deep Learning offers methods to control the synthesis process via learned high-level features and allows generating a wide variety of sounds. In this paper, we present \emph{DrumGAN VST}, a plugin for synthesizing drum sounds using a Generative Adversarial Network. DrumGAN VST operates on 44.1 kHz sample-rate audio, offers independent and continuous instrument class controls, and features an encoding neural network that maps sounds into the GAN's latent space, enabling resynthesis and manipulation of pre-existing drum sounds. We provide numerous sound examples and a demo of the proposed VST plugin.\footnote{\url{https://cslmusicteam.sony.fr/drumgan-vst/}}
\end{abstract}

\section{Introduction}
\label{sec:intro}

Drum sound design plays a prominent role in arranging and producing a song in most contemporary popular music. 
 Thanks to synthesizers and the availability of professionally-recorded sound libraries, producers can directly process drum samples in a computer or specialized hardware (e.g., in drum machines). However, the timbre diversity offered by sample packs or synthesizers is limited, and producers are forced to perform complex sound superpositions (i.e., layering) and envelope processing techniques
%to deviate from the available sounds.
to create variations. Interaction and exploration are other critical issues with such techniques, as browsing in sample packs or fiddling with expert controls in a synthesizer can be a barrier to creativity for some people. \looseness=-1

Fueled by recent advances in Deep Learning (DL), the field of neural audio synthesis has shown the potential to overcome some of the inconveniences mentioned above. Many deep generative models can learn high-level latent variables that provide more expressive and intuitive means for sound exploration and synthesis \cite{drumgan, drysdale2021sds, wavegan}. In addition, as DL models can be trained on arbitrary data, the sound diversity is not strictly limited to that of a particular synthesis process. Despite the unprecedented success of many of these methods in experimental settings, just a handful of works have culminated into production-ready tools for music creation (e.g., Mawf,\footnote{\url{https://mawf.io/}} Neurorack \cite{ninon}). The challenge of modeling high-quality raw audio at scale \cite{Dieleman2018}, coupled with the requirement for such music creation tools to operate in real-time in resource-limited environments, 
%is preventing their use in commercial applications. \looseness=-1
has remained the thorn in the side for their use in commercial applications. \looseness=-1

In this work, we present \textit{DrumGAN VST}, a drum sound synthesis plugin based on a prior work employing Generative Adversarial Networks (GANs) \cite{drumgan}. Driven by feedback from professional artists and music production standards, we perform a series of improvements on the original model: i) 44.1 kHz sample-rate audio operability; ii) continuous instrument control; iii) encoding-decoding of sounds to generate variations. Additionally, in Appendix~\ref{ap:vst}, we describe the development of a VST prototype and its integration into Backbone 1.5, by Steinberg Media Technologies.\footnote{\url{https://www.steinberg.net/vst-instruments/backbone/}}

The rest of the paper is organized as follows: In Sec.~\ref{sec:sota} we review prior work on neural drum sound synthesis, paying special attention to DrumGAN \cite{drumgan}; in Sec.~\ref{sec:contributions} we detail the proposed changes over the original model; Sec.~\ref{sec:exp} describes the experiments carried out. Results are presented in Sec.\ref{sec:res}, and we conclude in Sec.~\ref{sec:con}.

\section{Previous Work}\label{sec:sota}

Many deep learning methods have been applied to address general audio synthesis. \emph{Autoregressive models} are some of the most influential, achieving state-of-the-art results in many audio synthesis tasks, particularly for speech generation \cite{wavenet}. However, autoregressive methods are generally slow at generation and afford little control, essential in a music creation tool. \emph{Latent variable models} such as Generative Adversarial Networks (GANs) \cite{Goodfellow2016}, or Variational Auto-Encoders (VAEs) \cite{vae}, have been more widely used in this sense as they are faster and allow manipulating learned controls affecting high-level factors of variation in the generated data \cite{DBLP:journals/corr/abs-2111-05011, planetdrums, gansynth}. Specifically, GANs have shown promising results in drum sound synthesis \cite{drumgan, drysdale2021sds, DBLP:conf/mm/TomczakGH20} and are generally superior to other generative methods in terms of speed and quality. Recently, Denoising Diffusion models have shown results on par with GANs \cite{DBLP:conf/nips/HoJA20} and were applied to drum sound synthesis obtaining unprecedented audio quality and diversity \cite{DBLP:conf/ismir/RouardH21}. However, diffusion models require an iterative denoising process that is often time-consuming.
% (e.g., CRASH \cite{DBLP:conf/ismir/RouardH21} takes 7 seconds to generate 1 second of audio in a GPU, DrumGAN \cite{drumgan} takes 200ms in a CPU).

Our work builds upon DrumGAN \cite{drumgan} by carrying out a series of modifications aimed at building a functional drum synthesis tool complying with artists' workflows and industry standards. The following section briefly introduces the original architecture.

\subsection{DrumGAN}
\label{sec:drumgan}

DrumGAN \cite{drumgan} is a Generative Adversarial Network (GAN) trained to generate drum sounds conditioned on high-level perceptual features (e.g., \emph{boominess}, \emph{hardness}, \emph{roughness}). It operates on 16 kHz sample-rate audio in the form of a complex Short-Time Fourier Transform (STFT) spectrogram (i.e., using the \emph{real} and \emph{imaginary} components of the STFT as separate channels of a tensor). 

The input to DrumGAN's generator $G$ is a concatenation of $n_{C} = 7$ perceptual features $\boldsymbol{c}$ mentioned above, and, as it is typical in the GAN setting, a random vector $\boldsymbol{z}$ sampled from an independent Gaussian distribution \(\boldsymbol{z} \sim \mathcal{N}_{n_z = 128}(\mu = 0, \sigma^2=\mathbf{I})\) with $n_z=128$ latent dimensions. The resulting vector, with size $n_z + n_{C} = 135$, is fed to $G$ to generate the output signal \(G(\boldsymbol{z}, \boldsymbol{c})\). The discriminator \(D\) estimates the Wasserstein distance between the real and generated distributions \cite{Gulrajani2017}. Also, in order to encourage $G$ to use the conditional features, $D$ has to predict these and an auxiliary mean-squared error loss is added to the objective. 
The authors employ the \emph{progressive growing} framework \cite{Karras2017} where the architecture is built dynamically during training.
As for the architecture, $G$ and $D$ follow a mirrored configuration composed of a stack of $6$ convolutional blocks. Each block is followed by up/down-sampling steps (respectively for $G$ and $D$) of the temporal and frequency dimension. 
For more architecture and training details, we refer the reader to the original paper \cite{drumgan}.

\section{Contributions}\label{sec:contributions}
In this section, we describe in more detail our contributions to the original DrumGAN implementation. We depart from a preliminary prototype built upon the original model (details in Appx.~\ref{ap:vst}). The plugin was tested by 3 artists over a month, followed-up by informal discussion sessions to drive modifications (see Sec.~\ref{sec:sr}, \ref{sec:labels}, and \ref{sec:encoder}).

% \subsection{Prototype}\label{ref:prototype}
% DrumGAN \cite{drumgan} shows encouraging results for drum synthesis in terms of control, quality and diversity, yet, having it accessible through a command-line interface impedes artists to incorporate it into their music production workflow, which ultimately hinders the possibility to obtain valuable feedback. Hence, our first focus prior deriving a more elaborate music production tool from DrumGAN, is to develop a usable prototype featuring a graphical user interface that can be tested comfortably by artists. The Virtual Studio Technology (VST) standard\footnote{\url{https://developer.steinberg.help/display/VST/VST+Home}} for the integration of virtual effect processors and instruments into digital audio environments seems like the natural choice for this task. This standard is open-source and many C++ frameworks exist such as JUCE\footnote{\url{https://juce.com/}}, which allows embedding deep learning models in a straightforward way. In Fig~\ref{fig:vst1}, we show a simple interface developed with JUCE that naively exposes DrumGAN's parameters as simple sliders for the perceptual features, and a 2D pane for the manipulation of 2 components of the latent-noise (we use an additional button to swap across the 128 latent dimensions). Following the methodology described in prior work analysing the development of creative AI technology for music production \cite{ref:emmanuel}, we give this tool to three professional music producers to obtain feedback and guide improvements both on the interface and the model. Artists , and, ultimately, deploy a production-ready tool for drum sound synthesis. 

\subsection{Increasing the sampling rate}
\label{sec:sr}
One main concern of artists is DrumGAN's audio quality, especially for instruments with significant energy density at the high-end of the spectrum (e.g., snares, cymbals). Since DrumGAN generates 16 kHz sample-rate audio, not only the highest represented frequency is 8 kHz (i.e., the Nyquist frequency), but also, as a result, we argue that filtering artifacts derived from the model are more likely to appear in perceptually-relevant frequency regions \cite{DBLP:conf/icassp/PonsPCS21}. \emph{DrumGAN VST} is trained on drum sounds with a sample rate of 44.1 kHz (see Sec.~\ref{sec:data}). In order to exploit the same architecture without losing representational capacity due to the increased audio resolution, we halve the duration of the generated audio to 0.5 seconds instead of 1 (see Sec.~\ref{sec:data}).

\subsection{From perceptual features to soft class labels}
\label{sec:labels}

Another reiterated concern from artists was the inconsistent behaviour of DrumGAN's perceptual controls, which we opt to remove altogether, as well as the impossibility of directly choosing the specific instrument class to be generated (kick, snare, or cymbal). Since DrumGAN is not conditioned on class labels, the only way to achieve this is by jointly manipulating the conditional features and the latent noise. On the other hand, artists appreciate the possibility to perform interpolations across classes (e.g., continuously transforming a snare into a kick). To allow for such type of control in \emph{DrumGAN VST} we condition it on soft instrument labels, i.e., continuous instrument class probabilities instead of one-hot class vectors. This allows users to continuously and independently control instrument-specific features of the sound to be synthesized.
%  (i.e., some sort of "kickness", "snareness", or "cymbalness" control)
Soft labels are obtained by separately training a classifier of kick, snare, and cymbal sounds using an ad-hoc implementation of the Inception network architecture \cite{DBLP:conf/cvpr/SzegedyVISW16} trained on 128-bin Mel spectrograms. Similar to prior work \cite{darkgan}, we distill knowledge from this classifier into our generative model by predicting class probabilities on the training data and using these to condition the GAN as explained in Sec.~\ref{sec:drumgan} (with $n_C = 3$ now).\looseness=-1

\subsection{Adding an encoder}
\label{sec:encoder}
As mentioned in Sec.~\ref{sec:intro}, various techniques are exploited by producers to create variations of existing drum sounds (e.g., layering, ADSR manipulation). To allow creating variations in our tool without giving up high-level control, we separately train an encoder network that maps incoming sounds into the GAN's latent space, similarly to previous works~\cite{drysdale2021sds} (i.e., estimates the noise vector $\boldsymbol{z}$ and the instrument class probability $\boldsymbol{c}$). This way, one can generate variations of an  existing sound by simply encoding it and moving away from the initially predicted noise vector $\boldsymbol{z}$. \looseness=-1
%Further, by fixing $\boldsymbol{z}$ and changing the class probability $\boldsymbol{c}$, it is possible to transform some encoded sound into a different class while preserving some timbre properties.

% The training procedure is depicted in the diagram in Fig.~\ref{fig:encoder}. 
% First, a random batch of drum sounds is drawn from $G$. Then, the log-magnitude is computed and fed to the encoder.

The encoder $E$ is composed of a stack of 6 convolutional layers with channels in \{32, 64, 128, 128, 64, 32\}, kernels of size 3x3, alternating stride of 2x2 and 2x1, and padding of size 1x1. These layers are followed by 4 fully connected layers (see Appx.~\ref{ap:enc} for details). $E$ outputs an estimation of the latent parameters $\boldsymbol{z}$ and $\boldsymbol{c}$. The training objective is to minimize a reconstruction loss on the generation parameters $\boldsymbol{z}$ and $\boldsymbol{c}$, as well as the spectral distance between the original and reconstructed log-magnitude spectrograms generated from the predicted parameters. The resulting loss function is \\

\vspace{-1.1cm}
\begin{equation*}
    \label{eq:gan}
    \begin{split}
        L = \alpha \cdot \text{MSE}((\hat{\boldsymbol{z}},\hat{\boldsymbol{c}}), (\boldsymbol{z}, \boldsymbol{c})) + \beta \cdot \text{MSE}(G(\hat{\boldsymbol{z}},\hat{\boldsymbol{c}}), G(\boldsymbol{z}, \boldsymbol{c})),
    \end{split}
\end{equation*}

\vspace{-0.3cm}
where $(\hat{\boldsymbol{z}},\hat{\boldsymbol{c}}) = E(G(\boldsymbol{z},\boldsymbol{c}))$, $\text{MSE}$ denotes the Mean Squared Error, and $\alpha$ and $\beta$ are weighting coefficients for the latent vector and magnitude reconstruction errors respectively.

Despite \emph{DrumGAN VST}'s encouraging audio quality results (see Sec.~\ref{sec:res}), we observe some systematic bias in the form of inaudible, high-frequency artifacts. In our first attempt to train $E$, we notice that it takes advantage of this bias to encode samples into the latent space, over-fitting on the training data and failing to encode real drum sounds that don't exhibit these artifacts. Therefore, we threshold the spectrograms below $-1.5$ dB to remove silent parts, forcing the model to learn from more salient information.

\section{Experiment Setup}\label{sec:exp}
In this Section details are given about the experimental setup, including the dataset used and the evaluation method.

\subsection{Dataset}
\label{sec:data}
\emph{DrumGAN VST} is trained on a proprietary collection comprising over 300k one-shot audio samples equally distributed across kick, snare, and cymbal classes. Sounds have a sample rate of 44.1 kHz and variable lengths. Each sample is trimmed or zero-padded to a duration of 0.55 seconds as 80\% of the data is below this duration. We perform a 90\% / 10\% split of the data for validation purposes. As in DrumGAN \cite{drumgan}, the model is trained on the real and imaginary components of the Short-Time Fourier Transform (STFT). The STFT is computed using a window size of 2048 samples and 75\% overlapping. The generated spectrograms are inverted to the signal domain using the inverse STFT. \looseness=-1

The encoder $E$ is trained on a fixed set of latent random vectors $\textbf{z}$ and the corresponding generations $G(\textbf{z}, \textbf{c}$) (i.e., no real sounds are used to train $E$), where the soft class labels $\textbf{c}$ are obtained 
by running our classifier on randomly sampled instances of the dataset described above.
%from predictions of real instrument sounds randomly sampled from the dataset described above.

\subsection{Evaluation}\label{sec:eval}
We evaluate our tool by computing various objective metrics on the generated content. In the accompanying website\footnote{\url{https://cslmusicteam.sony.fr/drumgan-vst/}} we show extensive examples and musical material created by music producers using the tool. As suggested by prior work \cite{emmanuel}, we believe that having music released by artists is an indirect but critical way of validation for any creative music tool.

As for the objective evaluation, a common practice in the generative modeling literature is to measure the Inception Score (IS), Kernel Inception Distance (KID), and Fréchet Audio Distance (FAD).\footnote{\url{https://github.com/google-research/google-research/tree/master/frechet\_audio\_distance}} These metrics assess, to some degree, the quality and diversity of the GAN generations. In order to evaluate $E$, we also compute a set of audio reconstruction metrics: the Mean-Squared Error (MSE), Log-Spectral Distance (LSD), Signal-to-Noise Ratio (SNR), Distortion Index (DI), and the Objective Difference Grade (ODG). Note that DI and ODG are a computational approximation to users' subjective evaluations when comparing two signals.

\section{Results}
\label{sec:res}

\begin{table*}

\begin{minipage}{\columnwidth}

 \begin{center}
 \begin{tabular}{lllll}
  \toprule
   & $\uparrow$ IS & $\downarrow$ KID & $\downarrow$ FAD\\
  \midrule
   real data & 1.86 & 0.004 & 0.09\\
  \emph{DrumGAN VST}  & \textbf{1.83} & 0.009 & \textbf{1.49} \\
  \emph{Style-DrumSynth} & 1.64 & 0.085  & 1.72 \\
  \emph{CRASH} & 1.81 & \textbf{0.004} & 1.91 \\
  \bottomrule
 \end{tabular}
\end{center}
 \caption{Results of IS, KID, and FAD (see Sec.~\ref{sec:eval}), scored by \emph{DrumGAN VST}. We compare against real data and two baselines: \emph{Style-DrumSynth} \cite{drysdale2021sds} and \emph{CRASH} \cite{DBLP:conf/ismir/RouardH21}.}
 \label{tab:eval_gan}
\end{minipage}
\hfill
\begin{minipage}{\columnwidth}
\scriptsize
 \begin{center}
 \begin{tabular}{lllllll}
  \toprule
   & & $\uparrow$ DI & $\uparrow$ ODG & $\downarrow$ MSE & $\downarrow$ LSD & $\uparrow$SNR\\
   \midrule
  \multirow{3}{*}{DrumGAN VST} & \emph{gen}  & \textbf{0.14} & \textbf{-1.73} & 0.03 & \textbf{2.94} & \textbf{-1.67}\\
  & \emph{our}  & -0.06 & -1.92 & 0.06 & 7.36 & -3.17\\
  & \emph{sds} & 0.04 & -1.83 & 0.05 & 7.28 & -2.85\\
  & \emph{test} & -0.20 & -2.06 & \textbf{0.01} & 11.10 & -2.87\\
  \midrule
  \multirow{3}{*}{Style-DrumSynth} & \emph{gen}  & -1.12 & -2.78 & 0.03 & 10.05 & -2.88\\
  & \emph{our}   & -1.76 & -3.06 & 0.09 & 15.31 & -4.82\\
  & \emph{sds}  & -1.56 & -2.97 & 0.08 & 12.15 & -3.60\\
  & \emph{test}  & -2.27 & -3.21 & 0.02 & 26.18 & -6.98\\
  \bottomrule
 \end{tabular}
\end{center}
 \caption{Results of DI, ODG, MSE, LSD, and SNR (see Sec.~\ref{sec:eval}) computed on encoded and reconstructed pairs from i) generated data (\emph{gen}), ii) \emph{our} training data, iii) the baselines's training data (\emph{sds}), and iv) a \textit{test} set including unseen sounds (e.g., toms).}
 \label{tab:eval_enc}
  \end{minipage}
  \vspace{-0.5cm}
\end{table*}

% \begin{table}
% \small
%  \begin{center}
%  \begin{tabular}{lllll}
%   \toprule
%   & $\uparrow$ IS & $\downarrow$ KID & $\downarrow$ FAD\\
%   \midrule
%   real data  & 1.86 & 0.004 & 0.09\\
%   \emph{DrumGAN VST}  & \textbf{1.83} & 0.009 & \textbf{1.49} \\
%   \emph{Style-DrumSynth} & 1.64 & 0.085  & 1.72 \\
%   \emph{CRASH} & 1.81 & \textbf{0.004} & 1.91 \\
%   \bottomrule
%  \end{tabular}
% \end{center}
%  \caption{Results of IS, KID, and FAD (see Sec.~\ref{sec:eval}), scored by \emph{DrumGAN VST}. We compare against real data and two baselines: \emph{Style-DrumSynth} \cite{drysdale2021sds} and \emph{CRASH} \cite{DBLP:conf/ismir/RouardH21}.}
%  \label{tab:eval_gan}
% \end{table}

Table~\ref{tab:eval_gan} shows $G$'s evaluation results for the IS, KID, and FAD. We compare results against real data and two prior works: \emph{Style-DrumSynth} \cite{drysdale2021sds}, based on StyleGAN, and CRASH \cite{DBLP:conf/ismir/RouardH21}, based on denoising diffusion models.\footnote{We employ 100 denoising steps in the denoising process} Overall, DrumGAN scores the best results for most metrics, closely followed by CRASH.  \emph{DrumGAN VST} and \emph{CRASH} obtain an IS that is on a par with real data, which suggests that both models can generate diversity across the instrument classes and that the generated samples are somewhat classifiable into one of all possible classes. \emph{Style-DrumSynth} obtains slightly worse results, although this could be due to the mismatch between \emph{Style-DrumSynth}'s training dataset and the one used to train the Inception model. The KID reflects whether the generated data overall follows the distribution of real data in terms of timbre features (the Inception model is trained to predict instrument classes and features from the audio-commons timbre models\footnote{\url{https://github.com/AudioCommons/ac-audio-extractor}}). \emph{CRASH} obtains results that are on a par with real data, followed closely by \emph{DrumGAN VST}, suggesting that both models can generate sounds sharing timbral characteristics with real data.  \emph{Style-DrumSynth} obtains relatively worse KID, suggesting that the real and generated data diverge in terms of timbral features. Finally, FAD is a reference-free measure that correlates with the perceived audio quality of the individual sounds.
% (measuring co-variances within data instances). 
We observe that \emph{DrumGAN VST} obtains lower FAD than the baselines, suggesting that the generated audio contains fewer artifacts and resembles real data in terms of perceived quality.

% \begin{table}
% \scriptsize
%  \begin{center}
%  \begin{tabular}{lllllll}
%   \toprule
%   & & $\uparrow$ DI & $\uparrow$ ODG & $\downarrow$ MSE & $\downarrow$ LSD & $\uparrow$SNR\\
%   \midrule
%   \multirow{3}{*}{DrumGAN VST} & \emph{gen}  & \textbf{0.14} & \textbf{-1.73} & \textbf{0.03} & \textbf{2.94} & \textbf{-1.67}\\
%   & \emph{our}  & -0.06 & -1.92 & 0.06 & 7.36 & -3.17\\
%   & \emph{sds} & \textbf{0.04} & -1.83 & 0.05 & 7.28 & -2.85\\
%   & \emph{test} & \textbf{0.01} & -1.86 & 0.05 & 6.99 & -2.77\\
%   \midrule
%   \multirow{3}{*}{Style-Drumsynth} & \emph{gen}  & -1.12 & -2.78 & \textbf{0.03} & 10.05 & -2.88\\
%   & \emph{our}   & -1.76 & -3.06 & 0.09 & 15.31 & -4.82\\
%   & \emph{sds}  & -1.56 & -2.97 & 0.08 & 12.15 & -3.60\\
%   & \emph{test}  & -1.31 & -2.80 & 0.08 & 12.49 & -3.45\\
%   \bottomrule
%  \end{tabular}
% \end{center}
%  \caption{Results of DI, ODG, MSE, LSD, and SNR (see Sec.~\ref{sec:eval}) computed on encoded and reconstructed pairs from i) generated data (\emph{gen}), ii) \emph{our} training data, iii) the baselines's training data (\emph{sds}), and iv) a \textit{test} set.}
%  \label{tab:eval_enc}
%  \vspace{-0.5cm}
% \end{table}

In Table~\ref{tab:eval_enc} we present the evaluation results for the encoder $E$. Again, results are compared against \emph{Style-DrumSynth}, which also incorporates a separate encoder that maps sounds into the GAN's latent space, following a method analogous to ours. The metrics are computed on encoded/reconstructed pairs from different sets of data: \emph{gen} refers to generated data (i.e., each model encodes/reconstructs its own generated data), \emph{our} training data (see Sec.~\ref{sec:data}), \emph{Style-DrumSynth}'s training data (\emph{sds}), and, finally, a \emph{test} set containing percussive sounds including examples out of the training distribution (e.g., toms, shakers). Overall, \emph{DrumGAN VST} outperforms the baseline in most metrics. It is interesting to see that this is the case even for the baseline's training data (\emph{sds}), never seen by \emph{DrumGAN VST} at training. It is also surprising that \emph{DrumGAN VST} generally obtains better MSE and SNR performance than the baseline, considering that these metrics are highly sensitive to phase information and that \emph{DrumGAN VST}'s $E$ only receives as input the magnitude of the STFT. Nonetheless, SNR is negative for all models, indicating that the time-domain residual signal from the difference between encoded and decoded sounds has greater power than the actual encoded sound. Therefore, despite the superiority of \emph{DrumGAN VST} on this metric, we can argue that neither of the models does a good job in terms of phase preservation. However, in terms of magnitude spectrogram reconstruction, \emph{DrumGAN VST} seems to obtain a much better performance than the baseline as suggested by the lower LSD (which only compares log-magnitude spectrograms). Finally, \emph{DrumGAN VST} outperforms the baseline on the DI and ODG metrics, which are precise metrics used for the perceptual evaluation of the audio quality. ODG ranges from 0 to -4, where lower values denote greater quality degradation between signals. \emph{DrumGAN VST} obtains ODG values between -1 and -2, indicating that there exist slight impairments between encoded and decoded sounds, although, in the case of the baseline, these impairments are generally annoying ($\text{ODG} < -2$). Differences between \emph{DrumGAN VST} and the baseline are accentuated in the case of DI, which correlates to the ODG but has higher sensitivity towards poor signal qualities.

We conclude from these results that the proposed encoder $E$ can competently encode and decode sounds into \emph{DrumGAN VST}'s latent space, even for timbres never seen during training, and with better performance than \emph{Style-DrumSynth}. As we show in the website accompanying this paper, this is the case even for, e.g., vocal percussion (i.e., beat-box sounds).

% \subsection{Informal Evaluation}
% It has been noted before that good objective performance cannot be sufficient when assessing deep generative models for music creation, particularly in the context of contemporary popular music \cite{emmanuel}. Even when models are not objectively performing as researchers would wish due to e.g., systematic bias or artifacts, music producers could still find a creative interest in them. 

% \javier{comment on the quality, diversity, playability/usability, latent space meaningfulness, responsiveness of conditional information, ...}
% \subsubsection{Released Musical Content}
% \javier{Talk about successful released content (azzaro campaing, whim therapy, AI-DrumKit, ...}

% \vspace{-0.1cm}
\section{Conclusion}
\label{sec:con}
% \vspace{-0.1cm}

In this work, we presented \emph{DrumGAN VST}, a plugin for analysis/synthesis of drum sounds employing an autoencoding Generative Adversarial Network (GAN). The model operates on 44.1 kHz sample-rate audio, and it enables continuous control over kick, snare, and cymbal classes. When compared to prior work on neural synthesis of drums \cite{DBLP:conf/ismir/RouardH21, drysdale2021sds}, our model obtains better results according to objective metrics assessing the quality, diversity, and reconstruction of sounds. The proposed plugin is developed in collaboration with professional music artists from whom we show released musical material on an accompanying website. The tool is integrated into Backbone 1.5 by Steinberg.\footnote{\url{https://www.steinberg.net/vst-instruments/backbone/}}

\section*{Acknowledgements}
% Jake Drysdale,

We thank Jake Drysdale for assistance with \emph{Style-DrumSynth}, and for providing us with his training data for comparison with our model; it has greatly improved the quality of the manuscript.

\nocite{langley00}

\bibliography{example_paper}

\begin{thebibliography}{20}
\providecommand{\natexlab}[1]{#1}
\providecommand{\url}[1]{\texttt{#1}}
\expandafter\ifx\csname urlstyle\endcsname\relax
  \providecommand{\doi}[1]{doi: #1}\else
  \providecommand{\doi}{doi: \begingroup \urlstyle{rm}\Url}\fi

\bibitem[Aouameur et~al.(2019)Aouameur, Esling, and Hadjeres]{planetdrums}
Aouameur, C., Esling, P., and Hadjeres, G.
\newblock Neural drum machine: An interactive system for real-time synthesis of
  drum sounds.
\newblock In \emph{Proc. of the 10th International Conference on Computational
  Creativity, {ICCC}}, June 2019.

\bibitem[Caillon \& Esling(2021)Caillon and
  Esling]{DBLP:journals/corr/abs-2111-05011}
Caillon, A. and Esling, P.
\newblock {RAVE:} {A} variational autoencoder for fast and high-quality neural
  audio synthesis.
\newblock \emph{CoRR}, abs/2111.05011, 2021.
\newblock URL \url{https://arxiv.org/abs/2111.05011}.

\bibitem[Deruty et~al.(2022)Deruty, Grachten, Lattner, Nistal, and
  Aouameur]{emmanuel}
Deruty, E., Grachten, M., Lattner, S., Nistal, J., and Aouameur, C.
\newblock On the development and practice of ai technology for contemporary
  popular music production.
\newblock \emph{Transactions of the International Society for Music Information
  Retrieval}, 5\penalty0 (1), 2022.

\bibitem[Devis \& Esling(2021)Devis and Esling]{ninon}
Devis, N. and Esling, P.
\newblock {Neurorack: deep audio learning in hardware synthesizers}.
\newblock In \emph{{EPFL Workshop on Human factors in Digital Humanities}},
  Dec. 2021.

\bibitem[Dieleman et~al.(2018)Dieleman, van~den Oord, and
  Simonyan]{Dieleman2018}
Dieleman, S., van~den Oord, A., and Simonyan, K.
\newblock {The challenge of realistic music generation: modelling raw audio at
  scale}.
\newblock In \emph{{NeurIPS}}, pp.\  8000--8010, Montr{\'{e}}al, Canada, Dec.
  2018.

\bibitem[Donahue et~al.(2019)Donahue, McAuley, and Puckette]{wavegan}
Donahue, C., McAuley, J., and Puckette, M.
\newblock Adversarial audio synthesis.
\newblock In \emph{Proc. of the 7th International Conference on Learning
  Representations, {ICLR}}, May 2019.

\bibitem[Drysdale et~al.(2021)Drysdale, Tomczak, and Hockman]{drysdale2021sds}
Drysdale, J., Tomczak, M., and Hockman, J.
\newblock Style-based drum synthesis with gan inversion.
\newblock In \emph{Extended Abstracts for the Late-Breaking Demo Sessions of
  the 22nd International Society for Music Information Retrieval (ISMIR)
  Conference.}, 2021.

\bibitem[Engel et~al.(2019)Engel, Agrawal, Chen, Gulrajani, Donahue, and
  Roberts]{gansynth}
Engel, J., Agrawal, K.~K., Chen, S., Gulrajani, I., Donahue, C., and Roberts,
  A.
\newblock Gansynth: Adversarial neural audio synthesis.
\newblock In \emph{Proc. of the 7th International Conference on Learning
  Representations, {ICLR}}, May 2019.

\bibitem[Goodfellow(2017)]{Goodfellow2016}
Goodfellow, I.~J.
\newblock {{NIPS} 2016 Tutorial: Generative Adversarial Networks}.
\newblock \emph{CoRR}, abs/1701.00160, 2017.
\newblock URL \url{http://arxiv.org/abs/1701.00160}.

\bibitem[Gulrajani et~al.(2017)Gulrajani, Ahmed, Arjovsky, Dumoulin, and
  Courville]{Gulrajani2017}
Gulrajani, I., Ahmed, F., Arjovsky, M., Dumoulin, V., and Courville, A.~C.
\newblock Improved training of wasserstein gans.
\newblock In \emph{Proc. of the International Conference on Neural Information
  Processing Systems, {NIPS}}, Long Beach, CA, {USA}, Dec. 2017.

\bibitem[Ho et~al.(2020)Ho, Jain, and Abbeel]{DBLP:conf/nips/HoJA20}
Ho, J., Jain, A., and Abbeel, P.
\newblock Denoising diffusion probabilistic models.
\newblock In Larochelle, H., Ranzato, M., Hadsell, R., Balcan, M., and Lin, H.
  (eds.), \emph{Advances in Neural Information Processing Systems 33: Annual
  Conference on Neural Information Processing Systems 2020, NeurIPS 2020,
  December 6-12, 2020, virtual}, 2020.
\newblock URL
  \url{https://proceedings.neurips.cc/paper/2020/hash/4c5bcfec8584af0d967f1ab10179ca4b-Abstract.html}.

\bibitem[Karras et~al.(2018)Karras, Aila, Laine, and Lehtinen]{Karras2017}
Karras, T., Aila, T., Laine, S., and Lehtinen, J.
\newblock Progressive growing of gans for improved quality, stability, and
  variation.
\newblock In \emph{International Conference on Learning Representations,
  {ICLR}}, May 2018.

\bibitem[Kingma \& Welling(2014)Kingma and Welling]{vae}
Kingma, D.~P. and Welling, M.
\newblock Auto-encoding variational bayes.
\newblock In \emph{Proc. of the 2nd International Conference on Learning
  Representations, {ICLR}}, Banff, AB, Canada, Apr. 2014.

\bibitem[Nistal et~al.(2020)Nistal, Lattner, and Richard]{drumgan}
Nistal, J., Lattner, S., and Richard, G.
\newblock {DrumGAN: Synthesis of Drum Sounds With Timbral Feature Conditioning
  Using Generative Adversarial Networks}.
\newblock In \emph{Proc. of the 21st International Society for Music
  Information Retrieval, {ISMIR}}, Montréal, Canada, 2020.

\bibitem[Nistal et~al.(2021)Nistal, Lattner, and Richard]{darkgan}
Nistal, J., Lattner, S., and Richard, G.
\newblock Darkgan: Exploiting knowledge distillation for comprehensible audio
  synthesis with gans.
\newblock \emph{arXiv preprint arXiv:2108.01216}, 2021.

\bibitem[Pons et~al.(2021)Pons, Pascual, Cengarle, and
  Serr{\`{a}}]{DBLP:conf/icassp/PonsPCS21}
Pons, J., Pascual, S., Cengarle, G., and Serr{\`{a}}, J.
\newblock Upsampling artifacts in neural audio synthesis.
\newblock In \emph{{IEEE} International Conference on Acoustics, Speech and
  Signal Processing, {ICASSP}}, pp.\  3005--3009, Toronto, Canada, June 2021.
  {IEEE}.
\newblock \doi{10.1109/ICASSP39728.2021.9414913}.
\newblock URL \url{https://doi.org/10.1109/ICASSP39728.2021.9414913}.

\bibitem[Rouard \& Hadjeres(2021)Rouard and
  Hadjeres]{DBLP:conf/ismir/RouardH21}
Rouard, S. and Hadjeres, G.
\newblock {CRASH:} raw audio score-based generative modeling for controllable
  high-resolution drum sound synthesis.
\newblock In Lee, J.~H., Lerch, A., Duan, Z., Nam, J., Rao, P., van Kranenburg,
  P., and Srinivasamurthy, A. (eds.), \emph{Proceedings of the 22nd
  International Society for Music Information Retrieval Conference, {ISMIR}
  2021, Online, November 7-12, 2021}, pp.\  579--585, 2021.
\newblock URL \url{https://archives.ismir.net/ismir2021/paper/000072.pdf}.

\bibitem[Szegedy et~al.(2016)Szegedy, Vanhoucke, Ioffe, Shlens, and
  Wojna]{DBLP:conf/cvpr/SzegedyVISW16}
Szegedy, C., Vanhoucke, V., Ioffe, S., Shlens, J., and Wojna, Z.
\newblock {Rethinking the Inception Architecture for Computer Vision}.
\newblock In \emph{{IEEE} Conference on Computer Vision and Pattern
  Recognition, {CVPR}}, pp.\  2818--2826, Las Vegas, NV, USA, June 2016. {IEEE}
  Computer Society.
\newblock \doi{10.1109/CVPR.2016.308}.

\bibitem[Tomczak et~al.(2020)Tomczak, Goto, and
  Hockman]{DBLP:conf/mm/TomczakGH20}
Tomczak, M., Goto, M., and Hockman, J.
\newblock Drum synthesis and rhythmic transformation with adversarial
  autoencoders.
\newblock In Chen, C.~W., Cucchiara, R., Hua, X., Qi, G., Ricci, E., Zhang, Z.,
  and Zimmermann, R. (eds.), \emph{{MM} '20: The 28th {ACM} International
  Conference on Multimedia, Virtual Event / Seattle, WA, USA, October 12-16,
  2020}, pp.\  2427--2435. {ACM}, 2020.
\newblock \doi{10.1145/3394171.3413519}.
\newblock URL \url{https://doi.org/10.1145/3394171.3413519}.

\bibitem[van~den Oord et~al.(2016)van~den Oord, Dieleman, Zen, Simonyan,
  Vinyals, Graves, Kalchbrenner, Senior, and Kavukcuoglu]{wavenet}
van~den Oord, A., Dieleman, S., Zen, H., Simonyan, K., Vinyals, O., Graves, A.,
  Kalchbrenner, N., Senior, A.~W., and Kavukcuoglu, K.
\newblock Wavenet: {A} generative model for raw audio.
\newblock In \emph{Proc. of the 9th {ISCA} Speech Synthesis Workshop},
  Sunnyvale, CA, USA, Sept. 2016.

\end{thebibliography}
\bibliographystyle{icml2022}

%%%%%%%%%%%%%%%%%%%%%%%%%%%%%%%%%%%%%%%%%%%%%%%%%%%%%%%%%%%%%%%%%%%%%%%%%%%%%%%
%%%%%%%%%%%%%%%%%%%%%%%%%%%%%%%%%%%%%%%%%%%%%%%%%%%%%%%%%%%%%%%%%%%%%%%%%%%%%%%
% APPENDIX
%%%%%%%%%%%%%%%%%%%%%%%%%%%%%%%%%%%%%%%%%%%%%%%%%%%%%%%%%%%%%%%%%%%%%%%%%%%%%%%
%%%%%%%%%%%%%%%%%%%%%%%%%%%%%%%%%%%%%%%%%%%%%%%%%%%%%%%%%%%%%%%%%%%%%%%%%%%%%%%
\cleardoublepage
\appendix
\section{User Interfaces}
\label{ap:vst}
\begin{figure}[h]
\centering
\includegraphics[ width=\columnwidth]{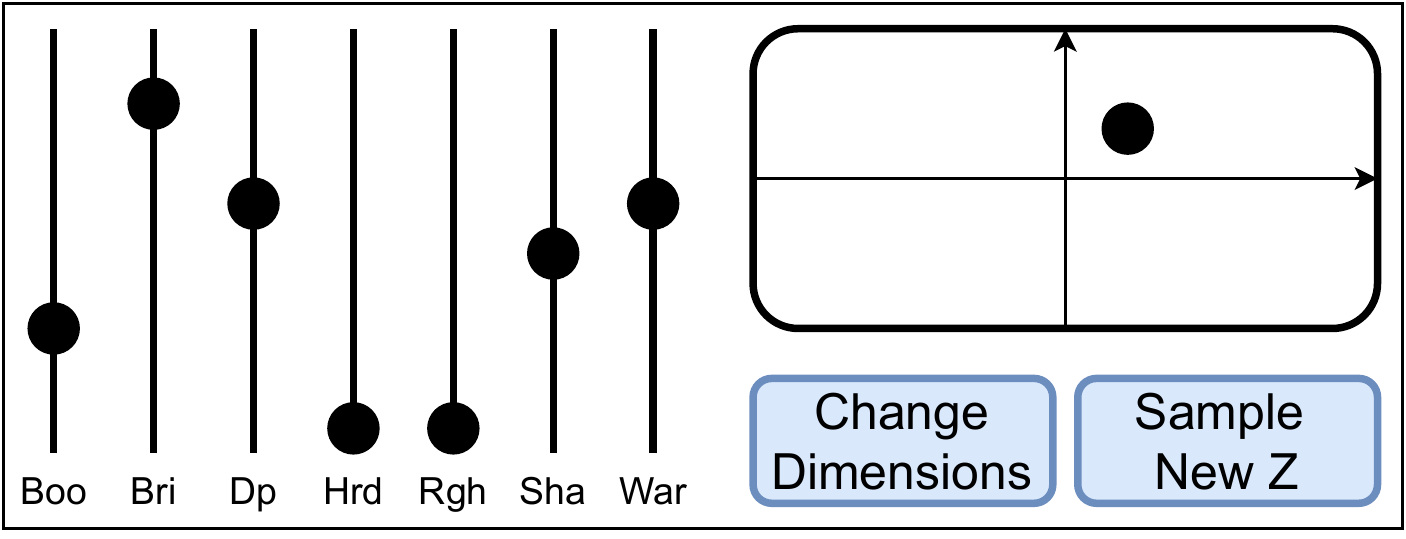}
\caption{Schematic of the first interface developed to interact with DrumGAN. It offers sliders to control the conditional perceptual features (e.g., \emph{boominess}, \emph{brightness}, \emph{depth}). A 2D plane, centered in $\boldsymbol{z}_{\text{center}}$ and directed by vectors $(\boldsymbol{e_1}, \boldsymbol{e_2})$ allows the user to explore different values for 
% $(z_{i\not{=} j}, z_{j\not{=}i})$ with $ j,i \in {1, ..., 128}$ dimensions from the random latent vector $\boldsymbol{z}\sim N_{128}(0, I)$
$\boldsymbol{z}$. $(\boldsymbol{e_1}, \boldsymbol{e_2})$ are orthonormal vectors sampled from a gaussian. A button (\emph{Change Dimensions}) allows to change these vectors and another button (\emph{Sample New Z}) allows to randomly sample a new center for the plane $\boldsymbol{z}_{\text{center}} \sim{N(0, I)}$. Ultimately, from the circled marker at coordinates $(\alpha, \beta)$ we decode $\boldsymbol{z} = \boldsymbol{z}_{\text{center}} + \alpha \boldsymbol{e_1} + \beta \boldsymbol{e_2}$}
\label{fig:prototype}
\end{figure}

% \begin{figure}[h]
% \centering
% \includegraphics[trim={0 7.5cm 6cm 2cm}, clip, width=\columnwidth]{figs/drumgan_GUI.png}
% \caption{Second interface of DrumGAN VST prototype. The perceptual controls are removed in favor of instrument-specific feature controls.}
% \label{fig:prototype2}
% \end{figure}

\begin{figure}[h]
\centering
\includegraphics[ width=\columnwidth]{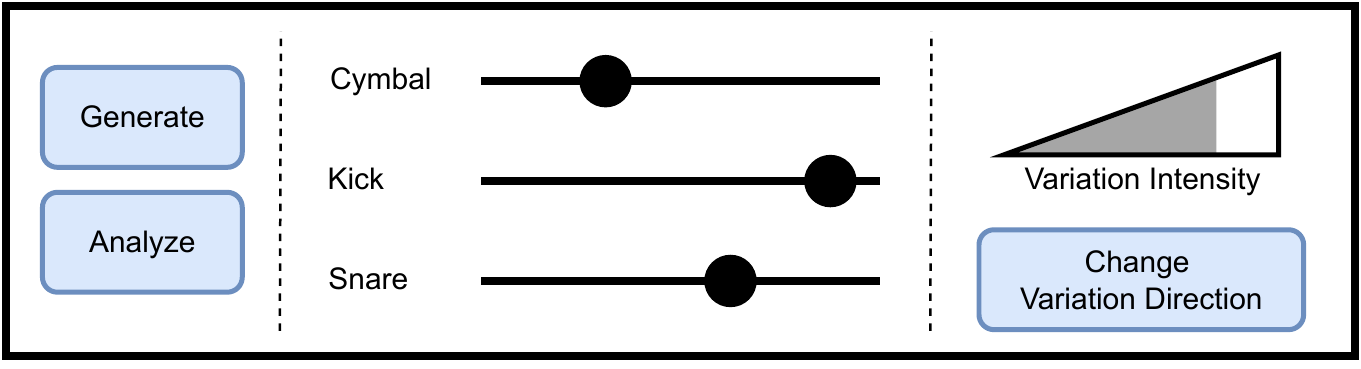}
\caption{Schematic of the interface developed for DrumGAN's integration into Steinberg's Backbone 1.5.
%Steinberg's BackBone. 
The interface now has sliders to control the soft-class conditional vectors $\boldsymbol{c}$. A button 'Analyze' triggers the encoding of a sample, setting the sliders to the appropriate values, as well as the internal variable $\boldsymbol{z}_{\text{center}}$. The button 'Generate' is used to sample a new $\boldsymbol{z}_{\text{center}}$. The 2D plane is replaced with a simpler 1D slider. From a  variation intensity $\alpha$, we decode $\boldsymbol{z} = \boldsymbol{z}_{\text{center}} + \alpha \boldsymbol{e_1}$}
\label{fig:vst1}
\end{figure}

DrumGAN \cite{drumgan} showed encouraging results for drum synthesis in terms of control, quality and diversity, yet, having it accessible through a command-line interface impedes artists from incorporating it into their music production workflow, which ultimately hinders the possibility of obtaining valuable feedback. Hence, our first focus prior to devising a more elaborate music production tool from DrumGAN is to develop a usable prototype featuring a graphical user interface that artists can test comfortably. The Virtual Studio Technology (VST) standard\footnote{\url{https://developer.steinberg.help/display/VST/VST+Home}} for the integration of virtual effect processors and instruments into digital audio environments seems like the natural choice for this task. This standard is open-source, and many C++ frameworks exist, such as JUCE\footnote{\url{https://juce.com/}}, which allows embedding deep learning models easily. In Fig~\ref{fig:prototype}, we show the schematic of a simple interface developed with JUCE that naively exposes DrumGAN's parameters: sliders are used for the perceptual features, and a 2D plane is used the traverse the latent space. 

This prototype is assessed by three professional music producers that provide feedback and guide improvements on the interface and the model. As a result of artist feedback, and in addition to the model modifications described in this work, we also conduct some improvements over our initial interface. Artists generally find it difficult to interpret the 2D plane used to control the latent space navigation. Hence, in the last interface version, depicted in Fig.~\ref{fig:vst1}, we replace this plane with a single slider (\emph{Variation Intensity}) that specifies the magnitude of the displacement from some initially sampled $\boldsymbol{z}$. The button \emph{Change Variation Direction} sets a new random direction for the displacement. This interface is finally chosen to be implemented into Backbone 1.5 version.\footnote{\url{https://www.youtube.com/watch?v=qLFo2udvWfA}}
% In this tool, \emph{DrumGAN VST} is exploited as a "smart" oscillator, where the subtractive synthesizer departs from GAN-generated sounds instead of pre-stored sounds from a sample library.
% \onecolumn
% \begin{figure}
%     \centering
%     \includegraphics[ width=\columnwidth]{figs/Captura de pantalla (8).png}
%     \caption{Caption}
%     \label{fig:backbone}
% \end{figure}

\section{Encoder Details}
\label{ap:enc}
\begin{table}[h]
\scriptsize
\begin{tabular}{llllllll}
Layer  & Type & Ch. & tensor size & ker. & str. & pad. & activation \\
\toprule
Output & FC   & -    & 131        & -           & -      & -       & SoftMax    \\
Layer9 & FC   & -    & 512        & -           & -      & -       & LReLU  \\
Layer8 & FC   & -    & 1024       & -           & -      & -       & LReLU  \\
Layer7 & FC   & -    & 3072       & -           & -      & -       & LReLU  \\
Layer6 & CNN & 32   & 16x6       & 3x3         & 2x1    & 1x1     & LReLU  \\
Layer5 & CNN & 64   & 32x6       & 3x3         & 2x2    & 1x1     & LReLU  \\
Layer4 & CNN & 128  & 64x12      & 3x3         & 2x1    & 1x1     & LReLU  \\
Layer3 & CNN & 128  & 128x12     & 3x3         & 2x2    & 1x1     & LReLU  \\
Layer2 & CNN & 64   & 256x24     & 3x3         & 2x1    & 1x1     & LReLU  \\
Layer1 & CNN & 32   & 512x24     & 3x3         & 2x2    & 1x1     & LReLU  \\
\midrule
Input     & -    & 1    & 1024x48    & -           & -      & -       & -       \\  
\bottomrule
\end{tabular}
\caption{Architecture details for the encoder. The tensor output by the last FC layer is split up in $\boldsymbol{z}$ and $\boldsymbol{c}$. The SoftMax is applied to $\boldsymbol{c}$ only, while $\boldsymbol{z}$ does not go through a non-linearity.}
\label{tab:enc}
\end{table}
%%%%%%%%%%%%%%%%%%%%%%%%%%%%%%%%%%%%%%%%%%%%%%%%%%%%%%%%%%%%%%%%%%%%%%%%%%%%%%%
%%%%%%%%%%%%%%%%%%%%%%%%%%%%%%%%%%%%%%%%%%%%%%%%%%%%%%%%%%%%%%%%%%%%%%%%%%%%%%%

The encoder \emph{E} architecture is detailed in Table~\ref{tab:enc}. Batch normalization is followed after every layer and bias are removed. We employ a learning rate of $10^{-4}$ and a batch size of $28$ training instances. We train the model using weights $\alpha = 1$, and $\beta = 3$ in loss $L$ (see Sec.~\ref{sec:encoder}) for the latent vector and magnitude reconstruction errors respectively.

\end{document}